# Anomalous interfacial dynamics of single proton charges in binary aqueous solutions

Jean Comtet*[1], Archith Rayabharam[2], Evgenii Glushkov[1], Miao Zhang[1], Avsar Ahmet[3], Kenji Watanabe[4], Takashi Taniguchi[4], Narayana R Aluru[2], Aleksandra Radenovic[1]

[1] Laboratory of Nanoscale Biology, Institute of Bioengineering, School of Engineering, École Polytechnique Fédérale de Lausanne (EPFL), Lausanne, Switzerland.
[2] Department of Mechanical Science and Engineering, University of Illinois at Urbana-Champaign, IL, USA
[3] Electrical Engineering Institute, École Polytechnique Fédérale de Lausanne (EPFL), Lausanne, Switzerland.
[4] Institute of Materials Science and Engineering, École Polytechnique Fédérale de Lausanne (EPFL), Lausanne, Switzerland.
[5] National Institute for Materials Science, Tsukuba, Japan.

* jean.comtet@gmail.com

**Understanding the dynamics of charge exchange between a solid surface and a liquid is fundamental to various situations, ranging from nanofiltration to catalysis and electrochemistry. Charge transfer is ultimately determined by physicochemical processes (surface group dissociation, ion adsorption, etc…) occurring in the few layers of molecules at the interface between the solid and the liquid. Unfortunately, these processes remain largely uncharted due to the experimental challenges in probing interfacial charge dynamics with sufficiently high spatial and temporal resolution. Here, we resolve at the single-charge scale, the dynamics of proton charges at the interface between an hBN crystal and binary mixtures of water and organic amphiphilic solvents (e.g. alcohol), evidencing a dramatic influence of solvation on interfacial dynamics. Our observations rely on the application of spectral Single Molecule Localization Microscopy (sSMLM) to two types of optically active defects at the hBN surface, which act as intrinsic optical markers for both surface protonation and interaction with apolar alkyl groups of the organic solvent. We use sSMLM to reveal interfacial proton charge transport as a succession of jumps between the titratable surface defects, mediated by the transport of the solvated proton charge along the solid/liquid interface. By changing the relative concentration of water in binary mixtures, we evidence a non-trivial effect on interfacial proton charge dynamics, leading at intermediate water concentration to an increased affinity of the proton charge to the solid surface, accompanied by an increased surface diffusivity. These measurements confirm the strong role of solvation on interfacial proton charge transport and establish the potential of single-molecule localization techniques to probe a wide range of dynamic processes at solid/liquid interfaces.**

Understanding the dynamics and transport of charges at solid/liquid interfaces is key to a number of physical, biological and chemical processes, ranging from biophysical transport[1,2] and nanofiltration[3,4], to energy harvesting[5,6], catalysis[7,8] and electrochemistry[9,10]. Interfacial charge dynamics is ultimately determined by physicochemical processes, such as dissociation of charged functional groups or specific adsorption of ions[11–14], which occurs in the few molecular layers between the solid surface and the solvent. While new insight has been obtained thanks to electrokinetic measurements[15,16], Second Harmonic Generation[17,18] dynamic Atomic

Force Microscopy[19,20] or time-resolved fluorescence[21,22], a fundamental understanding of the dynamics of these interfacial processes remains poor[23,24], due to the difficulties in obtaining surface-specific information at both high spatial and temporal resolution. In this context, we could recently resolve the diffusive dynamics of individual excess proton charge at the interface between defected hexagonal boron nitride and aqueous solutions[25], demonstrating the potential of single-molecule localization microscopy to probe single-charge dynamics at solid/liquid interfaces.

Importantly, the physicochemical properties of the solvent is expected to have a strong, yet poorly understood effect on interfacial charge dynamics. In aqueous media, one facile route to alter charge solvation and the hydrogen bonding structure is by mixing water with an organic amphiphilic solvent (e.g. acetone or alcohol such as methanol MeOH or ethanol EtOH) possessing both hydrophilic (polar –OH or =O) and hydrophobic (apolar alkyl -$CH_3$) moieties. These binary solutions show anomalous thermodynamic behavior in the bulk, due to incomplete mixing at the molecular scale[26–29]. Indeed, the presence of alkyl hydrophobic groups hinders the participation of these organic molecules to the water hydrogen-bonding networks[26] and modifies its topology, an effect which has been probed extensively through molecular dynamics simulations[27,30–33]. Because protons are transported in water along the H-bonding network, proton charge transport is also strongly affected by the presence of the organic solvent[34–36], an effect which is reinforced due to the amphiphilic caracter of the $H_3O^+$ ion itself[37]. The behavior of these mixtures is further complexified at interfaces, with reports of self assembly[38] and specific adsorption[39] at hydrophobic surfaces.

In this paper, we investigate at the single charge scale the complex relationship between solvation and proton charge dynamics at solid/liquid interfaces. We employ spectral Single Molecule Localization Microscopy at the interface between a defected hexagonal boron nitride (hBN) crystals and binary mixtures of water and organic solvents (MeOH, EtOH and acetone). By varying the relative amount of water in the mixtures, we fine-tune hydrogen-bonding in the liquid and probe how it affects proton charge dynamics at the hBN surface. We first show that spectral single-molecule localization microscopy (sSMLM) can serve as a chemically sensitive probe of the surface state, allowing to separate the photoluminescence signal due to the protonation of a first type of defect at the surface of the flake, from the adsorption of hydrophobic alkyl groups on a second defect type. sSMLM then allows us to reveal single excess proton trajectories as a succession of jumps between surface defects, mediated by the transport of the solvated proton charge along the solid/liquid interface. We evidence highly non-trivial dynamics of interfacial charges, characterized by an increased affinity of solvated charges with the solid surfaces at intermediate water concentrations, concomitant with an increase in interfacial diffusivity. Our measurements demonstrate the subtle role of solvation on interfacial proton charge dynamics and are rationalized through ab-initio molecular dynamics simulations.

**Spectral SMLM experimental set-up**

As shown in Fig. 1A, we use a spectral Single Molecule Localization Microscope (sSMLM) to probe the dynamics of optically active defects at the surface of hBN crystals in contact with various aqueous and organic solvents. The samples are multilayer boron nitride flakes, exfoliated from high-quality crystals[40]. These as-exfoliated hBN flakes host few intrinsic defect sites and we induce surface defects deterministically through a brief plasma treatment (*see SI*). Exciting the hBN crystal with a 561 nm (2.21 eV) laser leads to the selective excitation of surface defects with energies well within the 6 eV band gap of the material (Fig. 1A, green and red spots at the hBN surface), and a high NA oil-immersion objective collects the resulting photoluminescence signal from 580 nm to 700 nm (i.e. 2.14 to 1.77 eV).

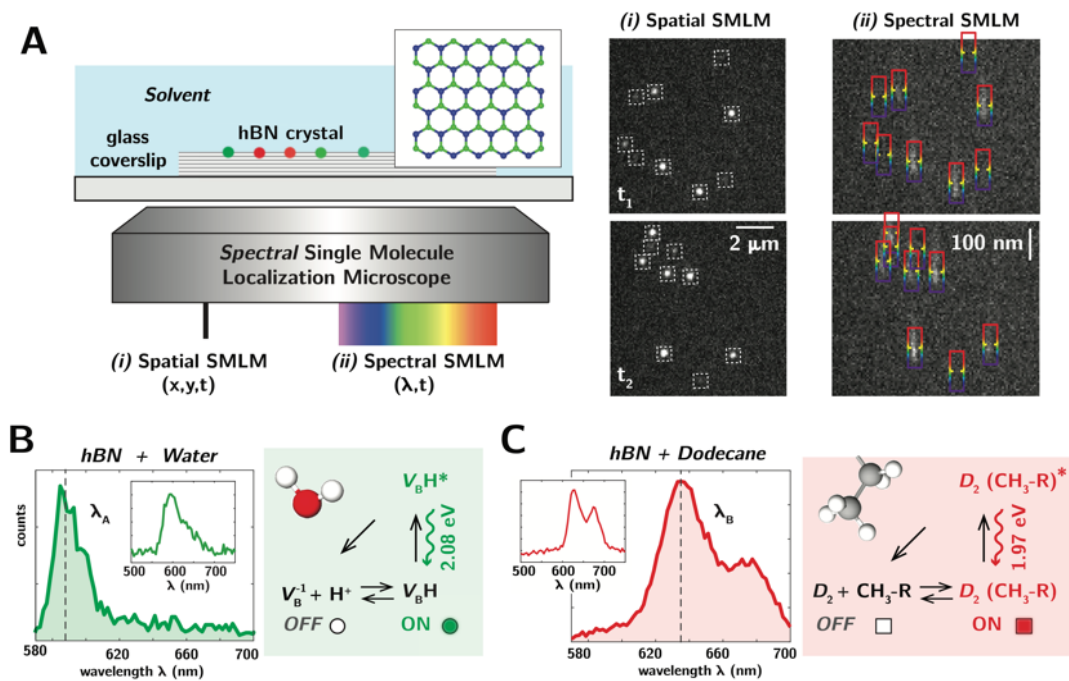

**Figure 1. Spectral super-resolution reveals chemical states of hBN surface defects. (A)** Spectral Single Molecule Localization Microscope (sSMLM) set-up, with an hBN crystal in contact with the solvent. Colored dots represent photoluminescence signal emanating from single defects at the surface of the flake. The inset shows the chemical structure of pristine hBN flake (boron in blue and nitrogen in green). The photoluminescence signal emitted from the flake's surface is split into spatial *(i)* and spectral *(ii)* channels. In the spatial channel *(i)*, emission from individual defects leads to diffraction-limited spots (highlighted by red boxes), localized with sub-pixel nanometric accuracy. In the spectral channel *(ii)* vertical dispersion by a prism allows the simultaneous measurement of the spectra of these individual emitters (highlighted by colored vertical lines). **(B-C)** The ensemble emission spectrum of hBN defects in contact with (B) water and (C) dodecane, showing respectively two main emission lines with $\lambda_A \approx$ 585 nm (2.08 eV) and $\lambda_B \approx$ 630 nm (1.97 eV). Representative spectra from individual emitters are shown in the inset. In **(B)**, emission $\lambda_A$ is due to of the protonation-induced transition between non-emissive deprotonated defect $V_B^-$ and emissive protonated defect $V_BH$, with excited state $V_BH^*$. In the excited state, $V_BH^*$ can either relax radiatively to its ground state (green arrow), or undergo excited state proton transfer and relax back to $V_B^-$ (black arrow). In **(C)**, emission $\lambda_B$ is due to interaction of the apolar hydrophobic alkyl group (-CH$_3$) with defect $D_2$ (see Fig. S1). Defect Zero Phonon Line is around 630 nm and the second peak visible around 670 nm corresponds to the phonon sideband.

Optically active emitters present at the flake's surface typically exhibit intermittent emission (blinking), leading to a sparse number of defects active on each frame. This sparse activation allows us to follow the dynamics of single defects using spectral Single Molecule Localization Microscopy (sSMLM) techniques. Our super-resolution set-up, shown in Fig. 1A and described previously[41,42], is augmented with the capability of spectral imaging, allowing to both localize individual defects with nanometric accuracy, while measuring their spectral response. Briefly, we collect the wide-field photoluminescence originating from the flake's surface and project half of the intensity on part of an EMCCD camera frame (*i*, spatial SMLM). In this spatial channel *(i)*, emission from individual defects leads to diffraction-limited spots on the camera, which we localize with subpixel accuracy, with a localization uncertainty $\sigma \approx \frac{\sigma_{\text{PSF}}}{\sqrt{N}} \approx 10 - 40$ nm, where $\sigma_{\text{PSF}} \approx 150$ nm is the standard deviation of the Gaussian fit of the emitter's intensity and $N$ the number of photons emitted by the defect during one acquisition frame. The other half of the photoluminescence signal is sent through a dispersive prism and is projected back onto the other part of the camera frame, allowing the simultaneous measurement of the emission spectra of the bright individual emitters (*ii,* spectral SMLM).

**Aqueous and organic solvents activate distinct types of defects**

Hosted at the surface of the flake, defects react strongly with their environment. As reported previously[25], a small density of active defects is observed in the air (typically $\rho \approx 10^{-3}$ defects.µm$^{-2}$.frame$^{-1}$, with 20 ms exposure time). For flakes in contact with the solvent, the dynamics and density of activated defects increase dramatically, pointing to the activation of defects by solvent molecules. Varying the solvent from aqueous to organic, we further observe that different solvents activate distinct types of defects. We compare in Figs. 1B-C the spectral response of flakes in contact with water, where free protons $H_3O^+$ are present in solution (Fig. 1B) and with dodecane, an organic hydrocarbon solvent (Fig. 1C), showing drastically different emission spectra depending on these two conditions, pointing to the activation of different types of defects.

For the case of hBN flake in contact with water (Fig. 1B), the ensemble spectra is characterized by an emission peak centered around $\lambda_A \approx 585$ nm (emission line A, green). Spectra from individual defects are shown in the inset. This emission line in hBN has been previously reported in a number of studies[25,41,43,44]. As represented in the green panel, we demonstrated in a previous report[25] that this emission line originates from defects in their protonated (acid) form, with the deprotonated (basic) defect being non-emissive. We attributed emission to a protonated boron vacancy $V_BH$, with the ON/OFF blinking behavior related to defect protonation/deprotonation through $V_B^- + H^+ \leftrightarrow V_BH$. Note that while our observations are fully consistent with an optical transition associated with disctinct protonation states, the chemical nature associated with this emission line is still under debate[45].

Consistent with the absence of free protons in the organic solvent dodecane, emission line A is absent on flakes in contact with this solvent (Fig. 1C). We observed instead the activation of emitters at a second emission wavelength, centered around $\lambda_B \approx 630$ nm (emission line B, red). This emission line has been reported previously in hBN[43,46–48] and corresponds to the activation of a second type of defects (of unknown chemical structure, which we denote as $D_2$). Due to the large increase in defect luminescence upon interaction with dodecane, we attribute the activation of defect luminescence as being due to interactions with its non-polar hydrophobic alkyl groups. Similar large increase of photoluminescence quantum yield and intensity in apolar environments have been reported in so-called solvatochromic dyes, and is attributed to interactions of the highly dipolar emitter excited state with the solvent[49]. Note that emission $\lambda_B$ is already activated by organic solvents even on pristine flakes (i.e. non-plasma treated) whereas emission $\lambda_A$ is greatly enhanced following plasma treatment, suggesting that these two emission lines are indeed due to defects with distinct chemical nature.

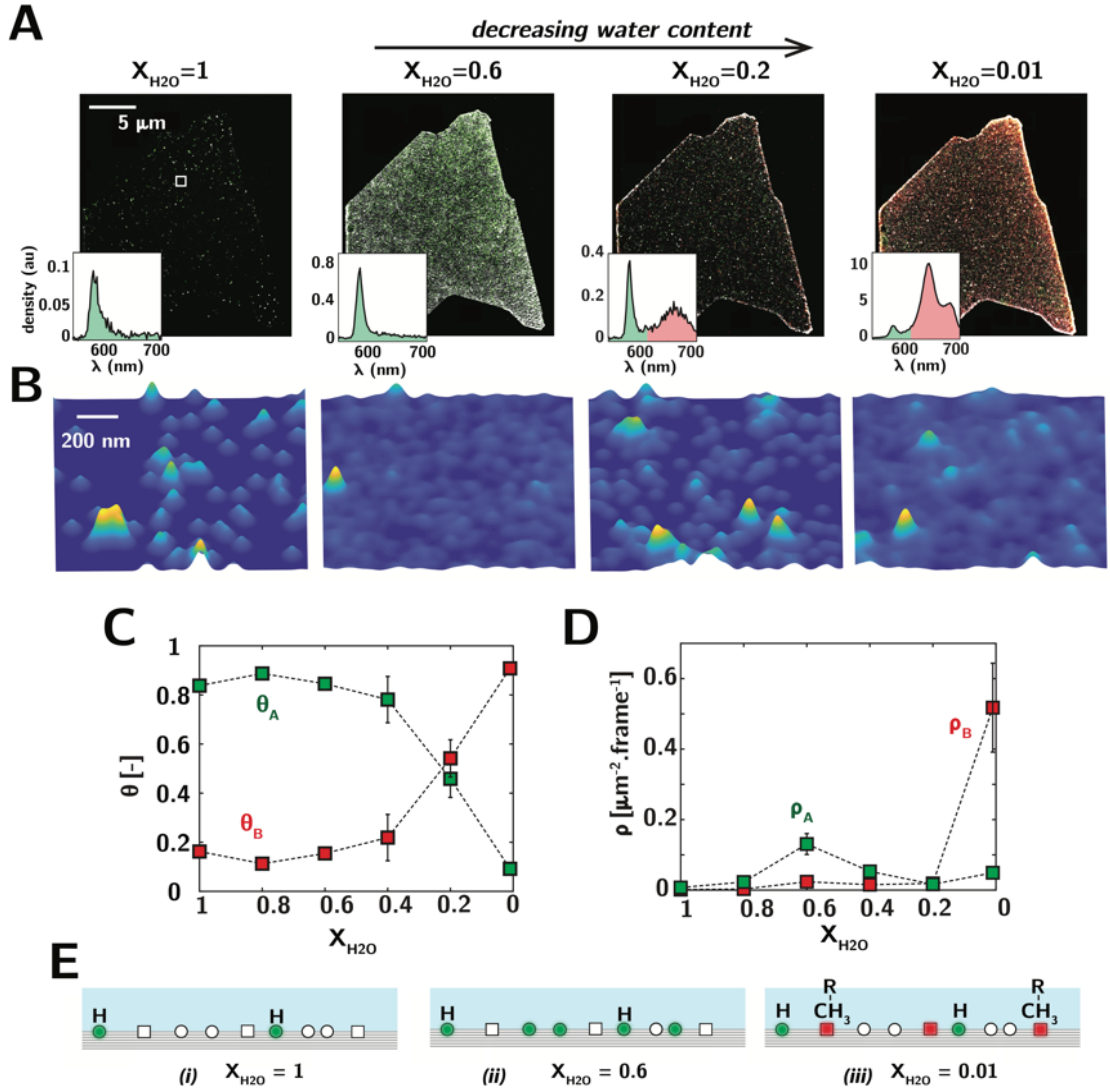

**Figure 2. Defect activity in binary water solutions. (A)** Reconstructed images of emitters on flakes in binary solutions of water and organic solvents (here water/acetone), at various volume fraction $X_{H2O}$ and with corresponding ensemble emission spectra in insets. Defects with emission spectra $\lambda_A$ and $\lambda_B$ are represented respectively in green and red. Defects with unassigned spectra are represented in white. The intensity scale for rendering defect density is the same in all conditions, except for $X_{H2O} = 0.01$, where the intensity scale is reduced by a factor of 10. **(B)** Zoom-in in the white box on the reconstructed images in (A), showing individual defect luminescence localization events, rendered with fixed uncertainty of 20 nm. **(C)** The relative proportion of the spectral population $\theta_A = N_A/(N_A + N_B)$ and $\theta_B = 1 - \theta_A$ for decreasing water content. **(D)** Respective density $\rho_A$ and $\rho_B$ of active defects per frame for decreasing water content. Error bars in (C) and (D) correspond to the standard deviation over two super-resolved images reconstructed from 10'000 frames. Larger error bars at $X_{H2O} = 0.4$ and 0.2 in (C) are due to the progressive increase of the number of alkylated defects over the two successive super-resolved images. **(E)** Schematic of defect surface state in various solvent conditions.

**Interfacial adsorption and defect activity in binary aqueous solutions**

Using the spectral signature of activated surface defects, we can probe how the physicochemical surface state of the hBN/liquid interface is affected by the presence of mixtures of water and amphiphilic organic solvents (MeOH, EtOH and acetone). We thus quantify the activity of defects under various proportions of water and organic solvents, characterized by the water volume fraction $X_{H2O}$ [-]. We show in Fig. 2A, colored super-resolved maps of active defects at the flake's surface, along with their ensemble spectra in the inset, and in Fig. 2B reconstructed local spatial density map of activated defects on a 1 μm² region at the surface of the flake (indistinctively of defect spectral signature). Quantifying defect activity and spectral emission allows us to get a direct readout of the local chemical and charge state of surface defects as a function of bulk solvent composition, varying systematically the proportion of water and amphiphilic organic solvents in contact with the flake. As shown on the ensemble spectra in the insets of Fig. 2A, we observe in all conditions the presence of the $\lambda_A = 585$ nm green emission line, characterizing the presence of protonated boron vacancies (Fig. 1D). At low water content ($X_{H2O} \leq 0.2$ in Fig. 2A) we observe the appearance of the second emission line $\lambda_B = 630$ nm, originating from the interaction of hydrophobic alkyl groups with surface defects $D_2$. While the dense and random distribution of defects present at the flake's surface prevents us from cross-corelating the spatial distribution associated with each defect type, achieving better control of defect locations and densities might allow for such quantitative correlations.

To further characterize the variation of interfacial defect state with bulk solvent composition, we plot in Fig. 2C the fraction $\theta_A$ and $\theta_B$ of protonated and alkylated defects characterized by each emission line for decreasing water molar fraction $X_{H2O}$, and in Fig. 2D the absolute density $\rho_A$ and $\rho_B$ of each type of defects. As quantified in Fig. 2C and schematically represented in Fig. 2D, the density of protonated boron vacancies $\rho_A$ first increases with decreasing water content, from $7.10^{-3}$ μm⁻².frame⁻¹.s⁻¹ in pure water (Fig. 2E, *i*) to a maximum of $130.10^{-3}$ μm⁻².frame⁻¹.s⁻¹ in 60% water (Fig. 2E, *ii*), pointing to an increased surface affinity of proton charges with the hBN surface in these mixed solvent conditions. Further decreasing water content below $X_{H2O} \leq 0.2$ leads to the appearance of a second defect type $D_2$ activated by the interaction with the hydrophobic alkyl groups (Figs. 2B-C, red, as schematically represented in Fig. 2D, *iii*). These observations are reported here for a mixture of water and acetone, but these trends are observed for the three organic solvents, with intrinsic variability when comparing individual flakes, due to local variation of the surface state (See *Fig. S3*).

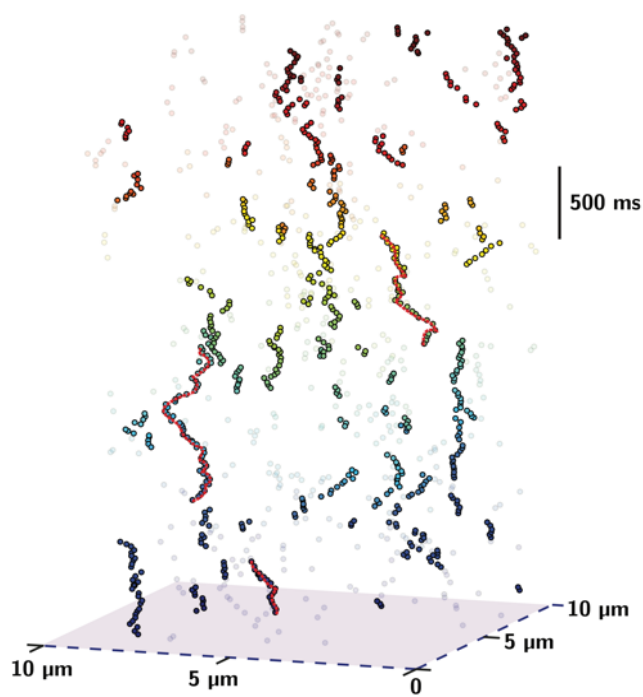

**Figure 3. Single-particle tracking of proton charge trajectories at the surface of hBN flake in binary solution** ($X_{H2O} = 0.6$). Spatiotemporal trajectories of single proton charges observed over 4 seconds, in a 10x10 μm² area over the hBN surface. The dimmer points correspond to uncorrelated blinking events. See *Movie S2*.

**Spatio-temporal correlations and single proton charge tracking**

Using the first type of defects as marker for protonation events (Fig. 1B), we can then track and follow the dynamics of excess protons moving at the solid/liquid interface with single charge resolution. We restrict our analysis to the steady-state regime emerging under continuous illumination, where the surface concentration of active defects is constant (See *Fig. S2*). Considering the condition $X_{H2O} = 0.6$, we reconstruct in Fig. 3 and movie *S2* a spatiotemporal plot of the activated defects in a 10x10 μm² area, with localized defects color-coded with increasing time along the vertical axis. Note that in this particular condition, only protonated defects are active at the surface of the flake (Fig. 2B), such that these trajectories are indeed solely due to the dynamics of proton charges.

As shown in Fig. 3, we see clear correlations in the activation of nearby defects over successive frames, and highlight some of these spatiotemporal trajectories in red. As defects are emissive in their protonated form, these trajectories correspond to the successive activation of nearby defects by a single excess proton hopping from defects to defects[25]. The fact that we can observe such trajectories demonstrates that the solvated proton charge, after desorbing from a defect, has a pronouced tendency to move along the solid/water interface, leading to the activation of a nearby defect site. This affinity of the proton charge to the hBN solid surface is thus at the root of these observed spatio-temporal correlations.

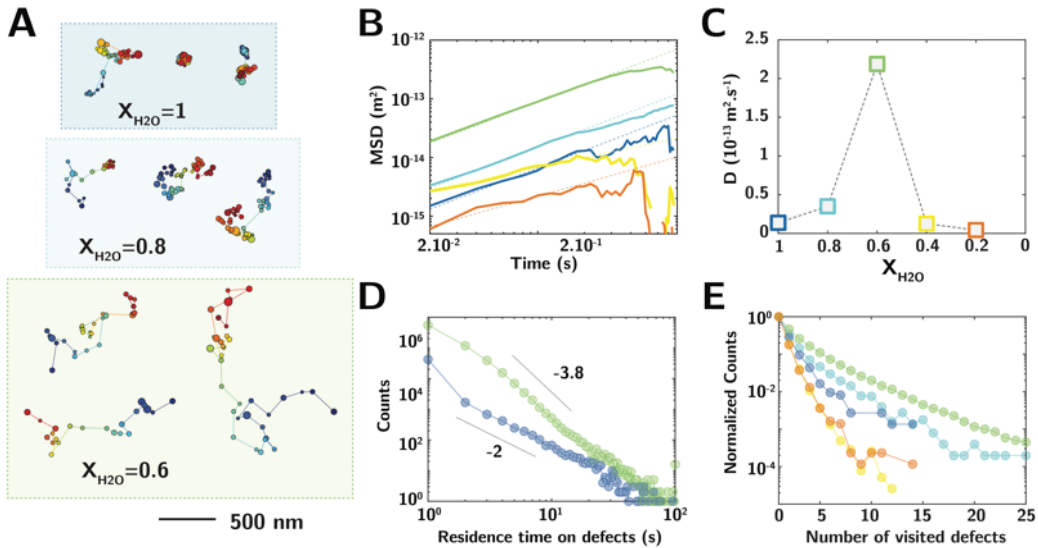

**Figure 4. Anomalous interfacial charge transport.** **(A)** Sample trajectories for various water volume fraction $X_{H2O}$ (in water/acetone mixture). **(B)** Evolution of the Mean Square Displacement (MSD) with time for various water volume fraction $X_{H2O}$. Green, ligth blue, dark blue, yellow and orange correspond respectively to $X_{H2O} = 1, 0.8, 0.6, 0.4$ and $0.2$. **(C)** Variation of the diffusion coefficient with water volume fraction $X_{H2O}$. **(D)** Distribution of the residence time on defects for $X_{H2O} = 1$ (blue) and $X_{H2O} = 0.6$ (green). **(E)** Distribution of the number of defect sites visited by individual trajectories for varying water fraction.

**Anomalous charge transport in mixtures**

Analyzing charge trajectories for various bulk solvent composition, we highlight in Fig. 4A representative trajectories observed at the flake's surface for the three water volume fraction $X_{H2O} = 1, 0.8$ and $0.6$, showing that the relative proportion of water and organic solvent has a dramatic effect on interfacial charge dynamics at the solid/liquid hBN interface. As we cannot attribute any spectra to low intensity defects (*Fig. S5*), we track and follow activated defects indistinctively of their emission spectra. Down to $X_{H2O} = 0.4$, these active emitters correspond solely to protonated defects (Fig. 2C, green), while for smaller water fraction, the localized emitters and the corresponding interfacial dynamics is caracteristic of both protonated and alkylated defects.

Focusing first on the conditions for which only protonated defects are present at the surface of the flake ($X_{H2O} \geq 0.4$), we plot in Fig. 4B the evolution of the Mean Square Displacement $MSD = <(x(t) - x(0))^2>$ over all observed trajectories $x(t)$ (See *SI*). From the initial increase of the $MSD$ with time, we extract a surface diffusion coefficient $D$ [m².s⁻¹] as $MSD \sim 4D.t$, which we report in Fig. 4C for decreasing water fraction. As shown in this figure, we observe a 20 fold increase in $D$, from $D \approx 10^{-14}$ m².s⁻¹ in pure aqueous solution to $D \approx 2.10^{-13}$ m².s⁻¹ at intermediate water fraction ($X_{H2O} = 0.6$). The diffusion coefficient further decreases when reaching lower fraction of water (note that in the conditions of low water fraction $X_{H2O} < 0.4$, the effective diffusion coefficient, and observed time traces, caracterizes interfacial mobility of both water and organic solvent molecules). This measured value $D \approx 10^{-13} - 10^{-14}$ m².s⁻¹ for the interfacial proton diffusion coefficient is 5 to 6 orders of magnitude lower than the bulk proton diffusion coefficient $D_{bulk} \approx 10^{-8}$ m².s⁻¹, consistent with a surface transport limited by proton charge desorption out of the defects[25]. Expressing the surface diffusion coefficient $D$, as $D \sim \frac{1}{4} k_{off}.a^2$ with $k_{off} = e^{-\Delta F/kT}$ [s⁻¹] the release rate of the proton out of the defect, $a$ [m] a characteristic inter-defect distance and $\Delta F$ [eV] a free energy desorption barrier, the 20-fold increase in interfacial mobility at intermediate water concentration would correspond to a decrease of the desorption barrier by 2.3 $k_B T$ or 60 meV.

To further investigate interfacial proton charge dynamics and probe in particular how the affinity of the solvated proton charge varies with water fraction, we analyse in detail the statistical properties of these observed random walks. As we are able to resolve individual trajectories, we can extract both the residence time at each defect sites (merging uncertainty-limited localizations as one site) and the number of visited defects along a single trajectory (See SI). We first show in Fig. 4D, the distribution of residence time on each defect site when the flake is in contact with aqueous solutions (blue, $X_{H2O} = 1$) and at intermediate water fraction (green, $X_{H2O} = 0.6$). These distributions follow power-law scaling (dashed lines with slope -2 and -3.8), and we observe a steeper distribution of residence time for $X_{H2O} = 0.6$ (comparing green and blue curves), consistent with the fact that the increase in surface diffusion observed at intermediate water concentration is due to the facilitated proton desorption out of the defects.

Finally, we plot in Fig. 4E the distribution of the number of defects visited during a single trajectory for various fraction of water. This number of visited defects is distributed approximately exponentially and follows the same non-monotonic trend as the diffusion coefficient, with a maximum in the number of visited defects at intermediate water fraction (Fig. 4E, green), consistent with the homogeneous and dense distribution of activated defects observed in this condition (Figs. 2A-B). Consistent with the large and homogeneous activation of defects observed at the flake's surface at intermediate water fraction, these observations indicate either an increased affinity of the proton charges to the solid surface, or an increased probability of readsorption to the surface defects. In particular, the microscopic affinity of the proton to the surface could be affected by the specific and anisotropic interaction of its hydrophobic oxygen side[35,37] with the alkyl group of the organic solvents.

**Simulation**

To probe mechanistically the kinetic factors affecting proton charge dynamics, we turn in Fig. 5 to ab-initio simulations of the reactivity of the negatively charged boron vacancies. We show in Fig. 5A the simulation cell, composed of a mixture of water and methanol molecules interacting with the hBN surface. Water/methanol mixture were considered for the simulations, as methanol represents the simplest organic chemical specie possessing an amphiphilic character due to its hydrophobic -$CH_3$ and its hydrophilic -OH, and is thus a good model system to understand the behavior of potentially more complex water/ethanol and water/acetone mixtures.

As shown in Fig. 5B, we compute the energy barrier for the transfer of proton from the $H_3O^+$ cation to the negatively charged boron vacancy $V_B^-$ (see SI). Varying, as shown in Fig. 5C, the first solvation shell of the hydronium from *(i)* two water, *(ii)* one water and one methanol, to *(iii)* two methanol molecules, we extract from these simulations the enthalpic desorption energy barrier for defects respectively in the ground state (blue, $\Delta E_{GS}$) and excited state (red, $\Delta E_{ES}$).

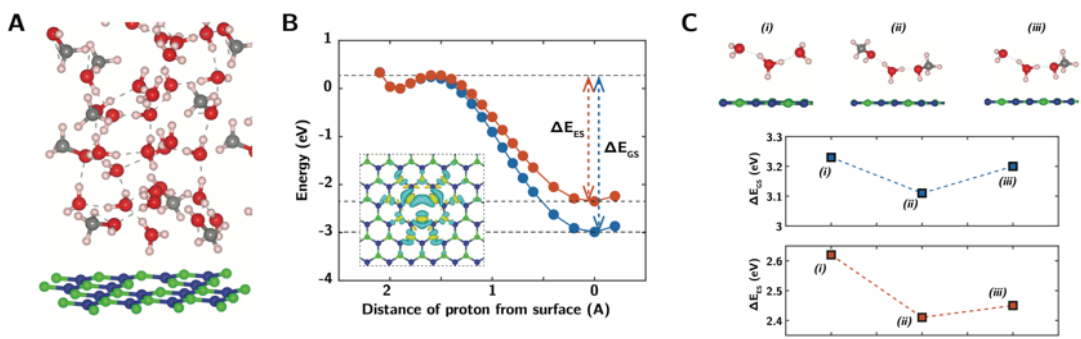

**Figure 5: Simulation of defect reactivity in binary mixtures. (A)** Simulation setup of a 50% water-methanol mixture on hBN. Oxygen is shown in red, hydrogen in white, carbon in teal, boron in brown and nitrogen in dark blue. The gray dashed lines represent the hydrogen bonds between oxygen and hydrogen atoms. **(B)** Energy barriers for a proton reacting with the negatively charged boron vacancy $V_B^-$ in a 50% mixture of methanol and water, with one water and one methanol molecule in the first solvation shells of the hydronium ion near the defect as shown in (A). Blue: defect in the ground state. Red : defect in the excited state. The desorption energy barrier is shown as $\Delta E$. The inset further shows the difference in the electron density between the excited and the ground state. Blue indicates a positive value of charge density difference and yellow indicates a negative charge density difference. **(C)** Desorption energy barrier $\Delta E_{GS}$ and $\Delta E_{ES}$ for a defect respectively in ground and excited state, with different molecules in the first solvation shell of the hydronium ion. *(i)* two water molecules *(ii)* one water and one methanol molecule *(iii)* two methanol molecules.

As reported in Fig. 5C, our simulations evidence a reduction of the enthalpic energy desorption barrier $\Delta E_{GS}$ and $\Delta E_{ES}$ by respectively ~120 meV and ~200 meV, when a single methanol molecule is present in the hydronium solvation shell, in qualitative agreement with the large increase in diffusion coefficient observed in Fig. 4C at intermediate water concentration. Note that the total free-energy barrier $\Delta F_{GS}$ for proton desorption will be futher reduced by a constant entropic contribution of -0.5 eV in the three conditions *(i-iii)* and might further decrease in mixture due to their anomalous mixing entropy[50] (see SI). On the contrary, the free-energy barrier for proton adsorption shows a slight increase of ~0.04 eV at intermediate water concentration (see SI), suggesting that the increase in the total residence of the proton at the surface (Fig. 4E) is indeed due to an enhanced surface affinity of the solvated charge, rather than an increased readsorption probability at the defect site.

Comparing desorption energy for defects respectively in the ground and excited state (Figs. 5B,C, comparing red and blue), we also find a 0.6 eV reduction of the desorption energy barrier for a defect in the excited state, consistent with our experimental observations that proton dissociation is favored under illumination (Fig. S2 and ref 25). The inset of Fig. 5B shows the difference in electron density between the excited and the ground state, evidencing a positive charge density difference around the defect in its excited state, consistent with a higher repulsion between the proton and the charge at the defect, leading to a global decrease of the desorption barrier. However, note that a quantitative agreement between the absolute computed desorption energy barrier and our experimentally measured diffusion coefficient is out of reach of these simulations. This could be due to either an incorrect defect type, with a less electronegative defect which would indeed decrease the desorption barrier, or to the fact that photoexcitation could lead to an excited state with higher energy, distinct from the one considered in this simulation.

**Conclusion**

Applying spectral Single Molecule Localization Microscopy to defected hBN crystals in contact with binary mixtures of water and organic solvents, we investigated at the single charge scale the complex relationship between solvation and charge dynamics at solid/liquid interfaces. We evidenced a non-trivial dynamics of interfacial proton charges, characterized by an increased affinity of solvated charges to the solid surface at intermediate water concentrations, concomitant with an increased interfacial diffusivity. Our measurements, corroborated by ab-initio simulations, demonstrate the subtle role of solvation on interfacial proton charge dynamics and further establish the potential of SMLM for the investigation of a wide range of dynamic processes at solid/liquid interfaces.

**Acknowledgement**
JC and AR thank Marie-Laure Bocquet, Benoit Grosjean, Michele Ceriotti and Mariana Rossi for discussion.

**Funding**
This work was financially supported by a Swiss National Science Foundation (SNSF) support through (200021_192037 grant) and National Centre of Competence in Research Bio-Inspired Materials.